\documentclass[twocolumn,runningheads,natbib]{svjour2}

\newcommand\ns {1E~1207.4$-$5209}
\newcommand\nss {1E1207}
\newcommand\snr {PKS~1209$-$51/52}

\newcommand\cxo {{\sl Chandra}}
\newcommand\xmm {{\sl XMM-Newton}}

\smartqed  
\usepackage{graphicx}

\journalname{Astrophysics and Space Science}
\begin{document}

\title{Evidence for a Binary Companion to the Central Compact Object
1E~1207.4$-$5209}


\author{
Peter M.\ Woods \and
Vyacheslav E.\ Zavlin \and
George G.\ Pavlov
}
\authorrunning{Woods, Zavlin \& Pavlov}


\institute{P.M. Woods \at
Dynetics, Inc. 1000 Explorer Blvd. Huntsville, AL 35806, USA \\
Tel.: +1-256-964-4914\\
\email{peter.woods@dynetics.com}
\and V.E. Zavlin \at
              Space Science Laboratory, NASA MSFC SD50, Huntsville,
AL 35805, USA \\
              Tel.: +1-256-961-7463\\
              \email{vyacheslav.zavlin@msfc.nasa.gov}           
\and G.G. Pavlov \at
Pennsylvania State University, 525 Davey Lab, University Park, PA 16802, USA \\
Tel.: +1-814-865-9482 \\
\email{pavlov@astro.psu.edu}
}

\date{Received: date / Accepted: date}

\maketitle

\begin{abstract}

Unique among neutron stars, \ns\ is an X-ray pulsar with a spin period of 424
ms that contains at least two strong absorption features in its energy
spectrum.  This neutron star is positionally coincident with the supernova
remnant \snr\ and has been identified as a member of the growing class of
radio-quiet  compact central objects in supernova remnants. From previous
observations with \cxo\ and \xmm, it has been found that the \ns\ is not
spinning down monotonically as  is common for young, isolated pulsars.  The
spin frequency history requires either  strong, frequent glitches, the presence
of a fall-back disk,  or a binary companion.  Here, we report on a sequence of 
seven \xmm\ observations of \ns\ performed during a 40 day window between  2005
June 22 and July 31.  Due to unanticipated variance in the phase measurements
during the observation period that  was beyond the statistical uncertainties,
we could not identify a unique phase-coherent timing solution.  The three most
probable timing solutions give frequency time derivatives of $+$0.9, $-$2.6,
and $+$1.6 $\times$ 10$^{-12}$ Hz s$^{-1}$ (listed in descending order of
significance).  We conclude that the local frequency derivative during our
\xmm\ observing campaign differs from the long-term spin-down rate by more than
an order of magnitude.  This measurement effectively rules out glitch models
for \ns.  If the long-term spin frequency variations are caused by timing
noise, the strength of the timing noise in \ns\ is much stronger than in other
pulsars  with similar period derivatives.   Therefore, it is highly unlikely
that the spin variations are caused by the same physical process that causes
timing noise in other isolated pulsars.  The most plausible scenario for the
observed spin irregularities is the presence of a binary companion  to \ns. We
identified a family of orbital solutions that are consistent with our
phase-connected timing solution, archival frequency measurements, and
constraints on the companions mass imposed by deep IR and optical observations.

\keywords{X-rays \and Neutron stars: individual: (\ns)
\and Supernovae: individual (PKS 1209$-$51/52)}
\end{abstract}



\section{Introduction}

There exist a handful of enigmatic X-ray point sources, very likely young
neutron stars, positionally coincident with supernova remnants (SNRs) whose
nature remains uncertain.  These objects are commonly referred to as central
compact objects (CCOs) that are characterized by soft, thermal X-ray  spectra
and an absence of ordinary pulsar activity such as radio pulsations,
$\gamma$-ray emission and pulsar wind nebulae (Pavlov et al.\ 2002a, 2004).

The CCO \ns\ (\nss\ hereafter) in the PKS 1209$-$51/52 SNR is a particularly
interesting member of this class in that it is both an X-ray pulsar (with a
0.424 s period; Zavlin et al.\ 2000)  and the only CCO found to possess
prominent  absorption lines in its  spectrum (Sanwal et al.\ 2002). The X-ray
energy spectrum is best modeled with a continuum blackbody component of
temperature $kT \approx$ 0.14 keV and at least two  broad absorption lines
centered at 0.7 and 1.4 keV. The strength of these lines depend upon the
rotational phase of the pulsar (Mereghetti et al.\ 2002).  Two additional
features, at 2.1 and 2.8 keV, have been reported (Bignami et al.\ 2003);
however,  their validity has been questioned (Mori, Chonco \& Hailey 2005).

The physical origin for the spectral lines remains unknown.   Sanwal et al.\
(2002) have concluded that these lines cannot be associated with transitions in
Hydrogen atoms and argued that neither electron nor proton cyclotron resonance
could cause these features. These authors suggest that the lines could be due
to absorption by once-ionized Helium in a magnetic field $B\sim 2\times
10^{14}$ G (see also Pavlov \& Bezchastnov 2005), while Hailey \& Mori (2002)
and Mori \& Hailey (2006) argue that the lines could be formed in an Oxygen
atmosphere with $B\sim 10^{11}$--$10^{12}$ G. As different interpretation imply
very different magnetic field strengths, measuring the field strength of \nss\
would be most important for understanding the nature of the spectral lines.  
For isolated pulsars, the most straightforward method for estimating  dipole
magnetic field strengths is to measure the spin frequency $\nu$ and its time
derivative (spin-down rate) $\dot{\nu}$.

Zavlin, Pavlov \& Sanwal  (2004) studied the spin evolution of \nss\ using a
compilation of \cxo\ and \xmm\ observations covering 3.5 years.  They found
that the spin frequency of this pulsar is not steadily decreasing as one would
expect for a magnetically-braking dipole.   Instead, the frequency evolution
was quite erratic, leading Zavlin et al.\ to consider three possible
explanations: ($i$) the star is undergoing frequent  glitches, ($ii$) the star
is surrounded by a debris disk that influences its spin evolution through
accretion and propeller torques, or ($iii$) the star is a member of a
(non-accreting) binary system.  Another possibility is that the spin-down of
the star is influenced by timing noise, a ubiquitous property of isolated
neutron stars  whose physical origin is unclear.  The sparse frequency history
of \nss\ could not distinguish between these models.

Here, we report on a sequence of seven \xmm\ observations of \nss\ that were
designed to perform phase-coherent timing in order to precisely measure the
pulse frequency and frequency derivative of the source.  A measurement of the
local frequency derivative  would  help us to distinguish between the possible
scenarios.  Below, we describe  the observation (\S2) and our timing analysis
of the \xmm\ data set (\S3),  and discuss the resulting constraints on the
physical mechanisms for the spin-frequency evolution in \nss\ (\S4).


\section{\xmm\ observations}

During a 40 day interval between 2005 June 22 and July 31,  \xmm\  observed
\nss\ seven times, with the EPIC PN camera as the primary instrument.   The
first three and last  two pointings had   effective exposures of 10$-$15 ks,
while the central (fourth) exposure was  about 45 ks.  The fifth pointing of
about 7 ks exposure was shorter than planned because of strong background
contamination.  The spacing between consecutive observations was approximately
15, 5, 1, 1, 5, and 15 days.   The spacing and durations of the \xmm\ pointings
were  planned in such a way as to allow phase-coherent timing of the pulsar
over the full time span of 40 days (see \S3). The exact exposures, observing
epochs, and other observational details for these observations are listed in
Table~\ref{tab:xmmobs}.


\begin{table*}[t]
\begin{center}
\caption{\xmm\ observation log for \ns.}
\label{tab:xmmobs}
\begin{tabular}{lllllllll}
\tableheadseprule\noalign{\smallskip}

\# &     \,\,\,\,ObsID      &     \,\,\,\,\,\,Date & Central epoch  &Span    & PN Exp.$^{a}$ & 
\,\,\,\,\,\,\,$r$$^{b}$ & Counts$^c$ & max $Z_{1}^2$
\\ 

&  &           & \,\,\,\,\,\,\,\,(MJD) & \,\,(ks)  &   \,\,\,\,\,\,(ks)            &   (arcsec)     & \\ 
\hline
 
\,1 &  0304531501    & 2005 Jun 22 & 53543.600542 & \,\,15.1  &   \,\,\,\,\,\,10.6            &      \,\,\,\,\,\,\,35 & \,\,13,233 & \,\,\,\,\,\,41.1 \\ 
\,2 &  0304531601    & 2005 Jul 05 & 53556.141927 & \,\,18.2 &   \,\,\,\,\,\,12.7            &      \,\,\,\,\,\,\,35 & \,\,12,858 & \,\,\,\,\,\,37.0 \\ 
\,3 &  0304531701    & 2005 Jul 10 & 53561.404084 & \,\,20.5  &   \,\,\,\,\,\,14.3            &      \,\,\,\,\,\,\,20 & \,\,17,651 & \,\,\,\,\,\,43.1 \\ 
\,4 &  0304531801    & 2005 Jul 11 & 53562.456311 & \,\,63.4 &   \,\,\,\,\,\,44.4            &      \,\,\,\,\,\,\,35 & \,\,56,804 &  \,\,\,120.4 \\ 
\,5 &  0304531901    & 2005 Jul 12 & 53563.335586 & \,\,\,\,\,9.6 &    \,\,\,\,\,\,\,\,\,6.7            &      \,\,\,\,\,\,\,20  & \,\,\,\,\,8,559 & \,\,\,\,\,\,19.1 \\ 
\,6 &  0304532001    & 2005 Jul 17 & 53568.112822 & \,\,16.5 &   \,\,\,\,\,\,11.5            &      \,\,\,\,\,\,\,35        & \,\,14,696 & \,\,\,\,\,\,81.2 \\ 
\,7 &  0304532101    & 2005 Jul 31 & 53582.691485 & \,\,17.7 &   \,\,\,\,\,\,12.4            &      \,\,\,\,\,\,\,20   & \,\,15,524 & \,\,\,\,\,\,26.5 \\ 
\hline
 & Sum & \,\,\,\,\,\,\,\,.... & \,\,\,\,\,\,\,\,\,\,\,\,\,\,\,\,....  & 161.0 & \,\,\,112.6 & \,\,\,\,\,\,\,\,.... & 139,325 & \,\,\,\,\,\,\,\,\,.... \\

\tableheadseprule\noalign{\smallskip}
\end{tabular}
\end{center}
\noindent$^{a}$ Effective source exposure times after filtering.  See text for details. \\
\noindent$^{b}$ Source extraction radius used for event selection. \\
\noindent$^{c}$ Number of counts used for timing analysis. \\
\vskip 0pt
\end{table*}

For each observation, the PN camera was operated in small window mode, with 5.6
ms time resolution.  Starting from the observation data files, all data were
processed using {\it XMMSAS} version 6.5.0.  After running the tool {\tt
epchain}, we extracted  light curves  from the observed field of view minus a
circular region that included \nss.   These light curves were used to identify
and filter out  periods of high background.   Source (plus background) counts
for timing analysis  were extracted from a circular region  for each
observation and filtered using standard criteria and the good time intervals we
determined.  The radii of the extraction regions, $35^{\prime \prime}$ or
$20^{\prime \prime}$, are listed in Table~\ref{tab:xmmobs}.  A smaller radius
of $20^{\prime \prime}$ was used to improve the signal-to-noise ratio for the
three  observations where the  background rate within the good time intervals
was elevated.  The filtered event lists were barycentered to the location 
${\rm R.A.} = 12^{\rm h} 10^{\rm m} 0.\!^{\rm s}80$,  ${\rm Decl.}=-52^\circ
26^{\prime} 25.\!^{\prime \prime}1$  using the {\it XMMSAS} tool {\tt barycen}. 
Finally, we selected counts within the energy range 0.4$-$2.5 keV to maximize
the signal-to-noise ratio of the pulsed signal before beginning our timing
analysis. The numbers of selected counts are given in Table~1 (background was
estimated to contribute less than 15\% in each dataset).

\section{Phase-Coherent Timing Analysis}

Phase-coherent timing analysis requires careful spacing of individual
observations such that an extrapolation of the measured phase model, 
$\phi(t)=\phi(t_0)+\nu (t-t_0)+\frac{1}{2}\dot\nu (t-t_0)^2+...$,
for a given
observation or set of observations is precise enough to predict the phase to
the next observation to much better than a pulse cycle.  The advantage of this
approach is that one can achieve far more precise measurements of the pulse
frequency and higher derivatives than by using independent pulse frequency
measurements with the same total exposure.  This approach is commonly applied
to all types of 
pulsars, including Anomalous X-ray Pulsars
(e.g.\ Gavriil \& Kaspi 2004), Soft Gamma Repeaters (e.g.\ Woods et al.\ 2002),
and radio-quiet Isolated Neutron Stars (Kaplan \& van Kerkwijk 2005a,b).

\subsection{Pulse Phase Fitting Technique}

As the frequency error ($\delta\nu$) in an individual observation is inversely
proportional to its duration,  $\delta\nu_j \propto T_j^{-1}$, the longer
central exposure served as our reference point.  We measured the pulse
frequency during this observation first via a  $Z^2_1$  search  (see \S 3.2)
and then refined this measurement as follows.  We split the observation into 4
segments and folded these segments on the measured frequency to generate pulse
profiles for each segment.  Next, we cross-correlated each pulse profile with a
high signal-to-noise pulse template and measured phase offsets.  The pulse
template is first derived from the central observation folded at the initial
frequency. The phase offsets for the 4 segments were fitted to a  straight
line  and the slope of this line was  added to the initial frequency  to
produce our  refined frequency.  The short gaps between the central exposure
and the  adjacent exposures were expected to preserve the phase information,
i.e. the  propagated phase error (e.g.,  between the 4-th and 5-th
observations, $\delta\phi = \delta\nu\,(t_5-t_4)$) was expected to be $\ll 1$
cycle, which would mean that  no pulse cycles are missed in the phase model. 
As one incorporates more and more data over a wider time span, the precision of
the phase model improves, and one can tolerate larger gaps between
observations.  Note that the template pulse profile is updated as more data are
included until the full data set is utilized.  By the time we incorporated the
measured phases from the first and final observations into  our  fit, it became
clear that the phase offsets did not conform to a simple linear trend, and a
quadratic term ($\propto\dot{\nu}$) was added to  the phase model, $\phi(t) =
\phi(t_0) + \nu (t-t_0) + \frac{1}{2} \dot{\nu} (t-t_0)^2$.  However, even the
inclusion of the quadratic term  did not reduce the variance of the phase
residuals to the point where we obtained an acceptable fit ($\chi^2$ = 19.2 for
8 degrees of freedom; see Fig.\ 1).

\begin{figure}
\centering
\includegraphics[width=3.0truein,angle=0]{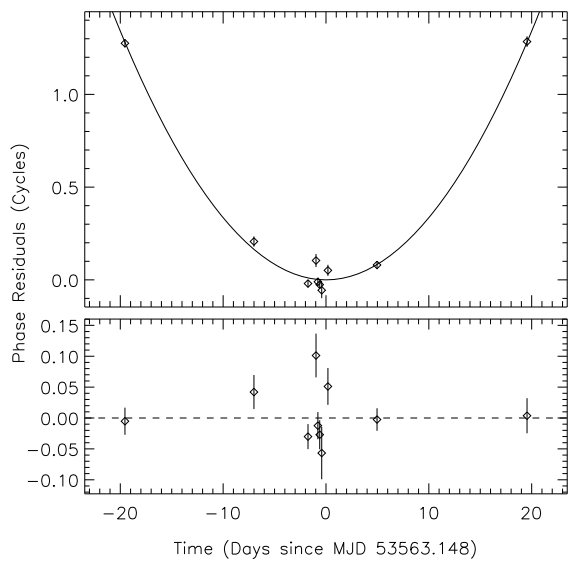}
\caption{Pulse phase residuals from \xmm\ observations of \nss\ during the 2005
observing campaign for model MOD1
(Table~2).  {\it Top}: Phase residuals minus a linear
trend.  {\it Bottom}: Phase residuals minus a quadratic trend.
Note that the central (longest) observation is split into four segments of
equal spans.
\label{fig:phases}}
\end{figure}

The poor fit to the quadratic phase model indicated that we either converged on
an alias solution or \nss\ exhibits significant ``phase noise''\footnote{In
this context, we refer to phase noise simply as deviations from our simple
quadratic phase model beyond statistical errors. Note that an alternative
definition of phase noise has specific meaning in the context of pulsar timing
noise (e.g.\ Cordes \& Helfand 1980).} on a time scale of weeks.  An alias
timing solution would be when there are an incorrect number of cycle counts
between consecutive observations.  Phase noise can be characterized in many
ways such as the presence of a strong  cubic term ($\propto\ddot{\nu}$),  white
noise, periodic variations, etc.  To ensure that the poor $\chi^2$ value in the
solution we found is {\em not} the consequence of misidentified cycle counts,
we employed a technique used for timing noisy rotators such as Soft Gamma
Repeaters (Woods et al.\ 2006).  In this technique, we measure the pulse phase
and frequency  at each of the 7 observing epochs. The phase for each
observation was measured by folding the data from each observation on a pulse
ephemeris of constant frequency determined by the central observation and
computing the phase difference between this profile and a template pulse
profile.  The pulse frequencies for the short observations were measured by
splitting the data into two segments of equal duration, folding each segment on
the pulse frequency measured for the central observation, measuring phase
shifts for each pulse profile relative to the pulse template, and fitting these
two pulse phases to a line to determine the local pulse frequency. Finally, we
perform a least-squares fit to the set of 7 phases and 7 frequencies, where we
vary the number of cycles between consecutive observations by integer
increments (Woods et al.\ 2006).  This provides a family of solutions to the
full data set which define the pulse phase evolution according to a quadratic
model covering the 40-day time interval.  None of the solutions (for a
quadratic phase model) provide a statistically acceptable fit to the data.  For
all possible solutions, the  ``null hypothesis probability'' (i.e. the
probability of measuring the large $\chi^2$ values by chance, assuming that the
model is correct) is very small.  Fit parameters for the top three solutions
(ranked in order of increasing $\chi^2$) are given in Table~\ref{tab:spin}.  We
chose to consider a limited number of solutions; therefore, we selected only
the solutions that had a probability of getting the measured $\chi^2$ by chance
of 10$^{-5}$ or larger (three solutions). We found that the best-fit model is
equivalent to the solution we identified via our bootstrap phase-fitting method
described earlier.  Phase residuals for this model are shown in
Figure~\ref{fig:phases}.  Since none of the identified solutions provide a
statistically-acceptable fit to the data, we conclude that \nss\ does, in fact,
exhibit significant phase noise on a time scale of weeks.   It seems unlikely 
that the excess noise we observed is due to underestimating our phase errors. 
Analysis of \xmm\ data from other pulsars using the same software, although
covering time spans shorter than 40 days, have consistently yielded reduced
$\chi^2$ values of $\sim$1 (e.g.\ Woods et al.\ 2004).


\begin{table*}[t]
\begin{center}
\caption{Candidate pulse ephemerides for \ns\ for 2005 June through July.
}
\label{tab:spin}
\begin{tabular}{llllllll}
\tableheadseprule\noalign{\smallskip}

Model & Epoch$^{a}$& $\nu^{b}$ & $\dot{\nu}$ & $\chi^2$/dof & Null Hypothesis & $Z_1^2$ & $Z_2^2$ \\
      &  (MJD TDB) &   (Hz)    & ($10^{-12}$ Hz s$^{-1}$) & & Probability \\

\hline
 
 MOD1 & 53563.148 & 2.357761722(16) &  $+$0.890(22)  & 25.2/11 & 2.9 $\times$ 10$^{-3}$ & 323.0 & 333.5 \\
 MOD2 & 53563.148 & 2.357762311(16) &  $-$2.651(23)  & 32.6/11 & 2.3 $\times$ 10$^{-4}$ & 314.9 & 325.8 \\
 MOD3 & 53563.148 & 2.357761720(17) &  $+$1.607(24)  & 37.6/11 & 3.6 $\times$ 10$^{-5}$ & 318.5 & 325.2 \\



\tableheadseprule\noalign{\smallskip}
\end{tabular}
\end{center}
\noindent$^{a}$ Pulse ephemerides are valid over the time range 
53543.547 to 53582.746 MJD TDB. \\
\noindent$^{b}$ Numbers given in parentheses indicate the 1$\sigma$ error in
the least  significant digit(s).  The statistical errors are inflated by a
factor 
$(\chi^2/{\rm dof})^{1/2}$.
\end{table*}

The presence of the phase noise  does not allow us to unambiguously
phase-connect the complete data set and thus measure a unique frequency and
frequency derivative for the full 40-day time span.   To place some constraints
on the frequency derivative during our observing sequence, we employed a
Monte-Carlo simulation to  estimate statistical significance of the multiple
solutions. For this simulation, we first had to choose a model for the phase
noise.  We selected two models: ($i$) a cubic phase term and ($ii$) white
noise.  In both cases, the amplitude of the model noise variance was equal to
the total variance in the top three fits to the data minus the statistical
variance. In our simulation, we generated phases for each observing epoch which
included three components: the model phase (including $\nu$ and $\dot{\nu}$
terms), Gaussian measurement noise, and the model phase noise.  In addition, we
simulated frequency measurements at each epoch assuming Gaussian measurement
noise (i.e.\ we neglect phase noise on the time scale of the observation
duration).  For each phase model, we generated 10$^5$ realizations and fit for
the cycle counts between consecutive epochs as we did for the measured data to
identify all possible timing solutions for each realization.  In each
realization, we identified the rank of the  true  timing solution in terms of
$\chi^2$.  The most constraining results were obtained from the white noise
model for the phase noise.  For this model, we found that the true timing
solution was among the top three solutions (ranked in order of $\chi^2$) 90\%
of the time and was the top solution 65\% of the time.  Assuming white phase
noise, our simulation suggests that we can be 90\% confident that the true
pulse ephemeris for \nss\ is MOD1, MOD2 or MOD3 given in Table~\ref{tab:spin}. 
Similarly, these results suggest there is a 65\% chance that MOD1 defines the
appropriate cycle counts between observing epochs, and hence, reflects the
correct pulse ephemeris.

The differences between the three pulse ephemerides listed in
Table~\ref{tab:spin} amount to small differences in the cycle counts between
the four outer observations in our observing sequence (i.e.\ a few additional
or  less cycles between observations 1 and 2, 2 and 3, 5 and 6,  and 6 and 7).
In fact, we can only be sure of the cycle count accuracy between the three
central observations (observations 3, 4 and 5 in Table~\ref{tab:xmmobs}).  To
show this explicitly, we fit for the cycle counts between the central three
observations as we did for the full data set, only we limited the order of  the
phase model to be first order on account of the short time span (2 days).  We
measure a difference in $\chi^2$ of 91 for 3 degrees of freedom between the
best-fit ephemeris identified in our search and the next closest.  Clearly, we
were able to phase-connect this subset of the data and unambiguously identify
the local pulse frequency ($\nu = 2.35776187(31)$ Hz over the time range
53561.328 to 53563.347 MJD TDB).

Although the method described in this section  is very efficient, it has some
limitations.  For large cycle count corrections between consecutive
observations, the local pulse ephemeris will change considerably as will the
folded pulse profile.  In turn, the pulse phase measurement will likely also be
affected.  In practice, the differences in the pulse shapes of \nss\ for the
three pulse ephemerides reported here are insignificant.  For very large cycle
count corrections, where this effect becomes important, the $\chi^2$
contribution from the frequency measurements begin to dominate the total
$\chi^2$, and these peaks are effectively suppressed. Even so, this method is
relatively new and not extensively tested.   To verify the results obtained
with this technique, we  employ the $Z_n^2$ test, a  traditional approach to
X-ray timing.

\subsection{The $Z^2_n$ test}

The $Z^2_n$ statistic (e.g., Buccheri et al.\ 1983) is defined as follows:
\begin{equation} Z^2_n = \frac{2}{N}\sum_{k=1}^{n}
\left[\left(\sum_{i=1}^N\cos\,2\pi k \phi_i\right)^2 +\left(\sum_{i=1}^N \sin\,
2\pi k \phi_i\right)^2\right], \end{equation} where $\phi_i =\nu (t_i-t_0) +
\dot{\nu} (t_i-t_0)^2/2 + ...$ is the phase of $i$-th event, $t_i-t_0$ is the
event arrival time counted from an epoch $t_0$ of zero phase, $n$ is the number
of harmonics involved in the test, and $N$ is the number of events.   For a
signal with a nearly sinusoidal pulse profile, such as observed from \nss, the
$Z_1^2$ (Rayleigh) test is known to give excellent results. For a sinusoidal
signal, the expected peak value of $Z_1^2$ is $N f_{\rm p}^2/2$, where $f_{\rm
p}$ is the pulsed fraction. For \nss, the pulsed fraction was measured to be
$f_{\rm p}=$8\%--12\% (Zavlin et al.\ 2000; Pavlov et al. 2002b).  The peak
$Z^2_{\rm 1}$ values found in the individual data sets  (Table~1) are in a
reasonable agreement with those given by  this estimate.

\begin{figure}
\centering
\includegraphics[width=3.5truein,angle=0]{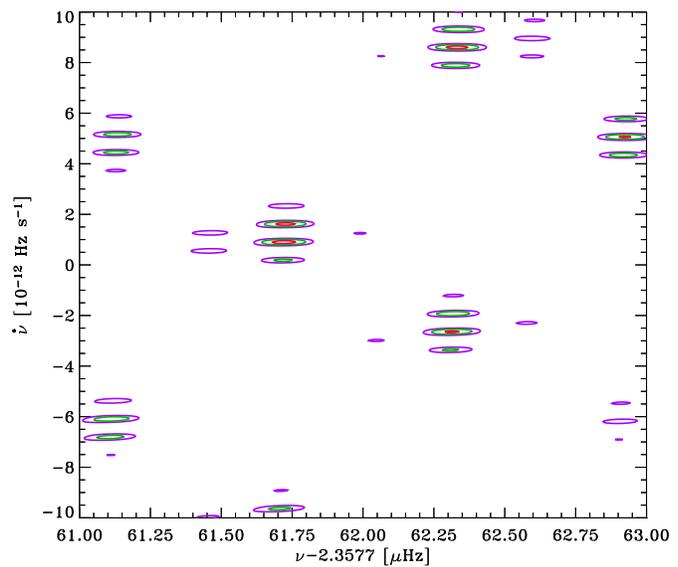}
\caption{
Contour plot of the $Z^2_1$ power on the $\nu$-$\dot{\nu}$ grid.
The purple, green and red contours correspond to
$Z^2_1=250$, 280, and 310, respectively.
\label{fig:grid}}
\end{figure}

The $Z_n^2$ test has been used for a phase-coherent timing analysis of several
observations spread over a large time span by Mattox et al.\ (1996) and Zavlin
et al.\ (1999), and we follow the approach described by those authors. To
account for the phase connection,  we apply the $Z_n^2$ test (for $n=1$ and 2)
to the whole data set of seven observations. To determine the parameters $\nu$
and $\dot\nu$ of the quadratic phase model, we calculated the $Z_n^2$  on a
dense two-dimensional grid  [$\nu-2.3577\,{\rm Hz}=41$--81 $\mu$Hz,
$|{\dot\nu}|<1\times 10^{-11}$ Hz s$^{-1}$], with $\nu$ and $\dot\nu$ spacings
of  0.02 $\mu$Hz and $2\times 10^{-14}$ Hz s$^{-1}$, respectively.   A contour
map obtained with the $Z_1^2$ statistic is shown in Figure~\ref{fig:grid}.
Because of the cycle-count ambiguities during the gaps between the consecutive
observations,  the map shows multiple peaks, one of them corresponding to the
true $\nu$,$\dot{\nu}$ solution and the others being aliases.  The  first,
third and fourth highest peaks  in this map correspond to MOD1, MOD3 and MOD2,
respectively (see Table 2). The top three peaks in a similar $Z_2^2$ map  are
at the same $\nu$,$\dot\nu$ as MOD1, MOD2 and MOD3, respectively. If the phase
connection between separate data sets were perfect, then the peak corresponding
to the true solution would be much higher than the aliases. However, in our
case the difference between the heights of the peaks turned out to be too small
to single out a unique solution. For instance, in addition to the highest peak
in the $Z_1^2$ map,  $Z^2_{1,{\rm max}}=323.0$ at $\nu=2,357,761.72$ $\mu$Hz,
$\dot\nu=+0.90\times 10^{-12}$ Hz s$^{-1}$, we see  four peaks with $310 <
Z^2_1< 320$ in Figure~\ref{fig:grid}, at different $\nu$,$\dot\nu$ values.
Similar to the method described in \S3.1, the differences in peak values of
$\nu$,$\dot\nu$ correspond to different (integer) numbers of cycles ($\sim
8\times 10^6$) during the full observational time span $T = 3393.8$ ks. We are
not aware of statistical criteria to estimate significance  of separate peaks
in this approach, and we can only assume that the solutions corresponding to
several highest $Z_n^2$ peaks cannot be ruled out on statistical grounds.  We
also note that  the lack of perfect phase-coherence is supported by the fact
that the largest $Z^2_{1}$ is much smaller than $\sum_{j=1}^7 Z_{1,j}^2 =
346.8$ (at the same $\nu$, $\dot{\nu}$),  the value we would expect to obtain
for perfect phase connection. Thus, the results of the  $Z_n^2$  search for the
\nss\ frequency and frequency derivative are  generally consistent with the
results reported in \S3.1.

\section{Origin of the erratic spin behavior}

The deviations from monotonic spin-down in \nss\ are substantial, and they
manifest on timescales of years to as short as possibly weeks as evidenced by
the phase noise detected here.   We now consider four possibilities for both
the erratic long-term spin behavior and short-term phase noise in \nss: ($i$)
frequent glitching, ($ii$) accretion and propeller torques from a circumstellar
debris disk, ($iii$) timing noise in an isolated neutron star, and ($iv$)
orbital Doppler shifts caused by the presence of a binary companion.

For any glitch model, the glitch frequency  and amplitude would have to be very
high  to account for the observed spin variability (see Zavlin et al.\ 2004).  
Moreover, the most viable timing solutions indicating spin down over the 40-day
observing span differ from the long-term spin-down of \nss\ over the last 5.5
years ($\sim -$4 $\times$ $10^{-14}$ Hz s$^{-1}$) by an order of magnitude. For
example, the most likely spin-down solution (MOD2) has a frequency derivative
more than one order of magnitude larger.   Because only a very contrived glitch
model could account for the long-term frequency history, and this model  would
provide no explanation for the short-term phase noise,  the glitch model is
effectively excluded by these observations.

Debris disks left over from the supernova explosions that produce neutron stars
could alter the spin evolution of the central neutron star via accretion and
propeller torques (Zavlin et al.\ 2004).  If the spin-up rate of \nss\ during
the 40-day interval were equal to the values measured for MOD1 or MOD3, then
the mass accretion rate would have to be very large ($\dot{m} >  3\times
10^{16}$ g s$^{-1}$).  Such a large accretion rate would require a large
increase in X-ray luminosity which is not observed.  Even the spin-down
solutions would require significant optical and IR emission from the disk. 
Deep IR observations of \nss\ have shown no indication of even a cool, passive
debris disk (Wang, Kaplan \& Chakrabarty 2006).  Thus, it appears unlikely that
a debris disk is the cause of the spin variability in \nss.

\begin{figure}
\centering
\includegraphics[width=3.3truein,angle=0]{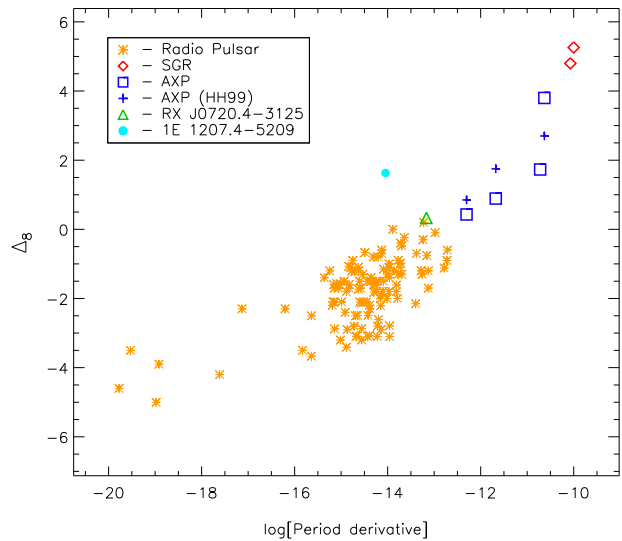}
\caption{The timing noise parameter $\Delta_8$ after Arzoumanian et al.\ (1994)
for 119 radio pulsars, the isolated neutron star RX~J0720.4$-$3125 (Kaplan \&
van Kerkwijk 2005b), four Anomalous X-ray Pulsars (blue squares - Gavriil \&
Kaspi 2004; blue plus signs - Heyl \& Hernquist 1999), two Soft Gamma Repeaters
(Woods et al.\ 2002) and the CCO 1E~1207.4$-$5209.
\label{fig:delta8}}
\end{figure}

Timing noise (irregular evolution of the pulse phase with time) is a ubiquitous
phenomenon in isolated neutron stars. This variability is in addition to the
usual variation caused by magnetic braking. It has been demonstrated that the
magnitude of these irregular variations depends upon the spin-down rate of the
pulsar  (e.g., Cordes \& Helfand 1980).  Millisecond pulsars show the smallest
timing noise while magnetars exhibit very strong timing noise.  In the case of
magnetars, these variations manifest as changes in the effective spin-down rate
of up to factors of 5 on a time scale of years.  For a convenient (albeit
crude) description of timing noise, Arzoumanian et al.\ (1994) introduced a
``stability parameter'' defined by the following equation: $\Delta_{\log t}  =
\log (|\ddot{\nu}|t^3/6\nu)$, where $t$ is the time during which the pulse
phase has been monitored ($t = 10^8$ s is a commonly used characteristic time),
and $\ddot{\nu}$ is the formal value of the second frequency derivative
obtained from fitting a cubic model to pulse phases (it is much larger in
magnitude than the actual $\ddot{\nu}$  for noisy pulsars).  Third-order
polynomial fits to the \nss\ phase residuals of the top three candidate timing
solutions yielded insignificant measurements of $\ddot{\nu}$. The timing
observations of \nss\ during 5.5 years were too sparse to fit the pulse phases
with any model. Therefore, to estimate $\ddot{\nu}$ and $\Delta_8$,  we fitted
the dependence $\nu(t) = \nu_0 + \dot{\nu}(t-t_0) + \ddot{\nu} (t-t_0)^2/2$ to
the frequency history covering the last 5.5 years. (The same exercise was
performed by Heyl \& Hernquist 1999 for some Anomalous X-ray Pulsars, and it
was shown to be reasonably accurate by Gavriil \& Kaspi 2004.)    Choosing $t_0
= 52700$ MJD TDB, we found $\nu_0= 2,357,762.8\pm0.2$ $\mu$Hz, $\dot{\nu} =
(-3.4\pm0.6) \times 10^{-14}$ Hz s$^{-1}$, and $\ddot{\nu} = (5.9\pm1.9) \times
10^{-22}$ Hz s$^{-2}$, which translates to a timing noise level of $\Delta_8 =
1.6$. In Figure~\ref{fig:delta8}, we show the period derivative versus the
timing noise parameter $\Delta_8$ for 126 isolated pulsars of various flavors
as well as for the CCO \nss.  The isolated pulsars fall along a relatively
well-defined locus, from the quiet millisecond pulsars to the noisy Soft Gamma
Repeaters, while \nss\ stands out from this trend with an anomalously large
timing noise strength, some 2$-$4 orders of magnitude higher than isolated
pulsars at similar spin-down rates.  Although such an estimate for the timing
noise parameter is, by necessity, very crude, its enormously high magnitude,
together with the gross inconsistency of the local (June-July 2005) spin-down
rate  with the long-term average, suggest that the erratic frequency behavior
in this source is not due to the same effect  that causes timing noise in other
isolated neutron stars.

\begin{figure}
\centering
\includegraphics[width=3.3truein,angle=0]{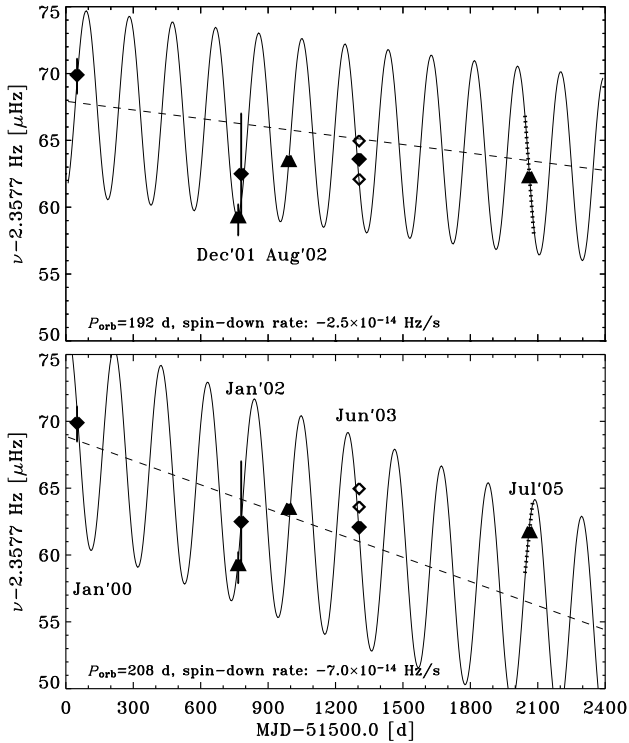}
\caption{Candidate orbital solutions for \nss\ consistent with our MOD2 (top)
timing solution and MOD1 (bottom) timing solution.
\label{fig:orbit}}
\end{figure}

The most straightforward explanation for the long-term spin variations in \nss\
is the presence of a binary companion.  Note that this model cannot explain the
observed short-term phase noise.  Current IR and optical limits for \nss\
exclude main sequence companions earlier than M5 and even white dwarfs with
effective temperatures greater than $\sim$10$^4$ K  (Fesen et al.\ 2006;  Wang
et al.\ 2006). Allowable companion masses are less than 0.2 M$_{\odot}$ for
late-type stars (Moody et al.\ 2006,  in preparation). Even with such  
low-mass companions, the resulting Doppler shifts are large enough to account
for the frequency variations in \nss.  Using the archival spin frequencies in
combination with the phases from our three candidate timing solutions, we fit
the data to a circular orbital model whose phase evolution is defined by the
following equation: $\phi(t) = \phi(t_0) + \nu (t-t_0) + \frac{1}{2} \dot{\nu}
(t-t_0)^2 + A \sin{\omega (t-t_0)}$.  This is the same equation as given in
\S3.1 with an additional sinusoidal term to account for the orbital Doppler
shifts.  We identified a family of acceptable orbits for each of the three
timing solutions listed in Table~\ref{tab:spin}.  The full set of allowable
timing solutions are too numerous to list.  We can place only very crude
constraints on the orbital periods to fall between  120 and 600 days. The mass
functions range between $1 \times 10^{-7}$ and $5 \times 10^{-5}$ M$_{\odot}$
for acceptable orbital solutions.  For a 90$^{\circ}$ inclination and a 1.4
M$_{\odot}$ neutron star, the corresponding companion mass range is 0.007 to
0.05 M$_{\odot}$, well within the existing limits on companion masses from IR
and optical observations.  For illustrative purposes, we show two example
orbital solutions that are consistent with the existing timing data for \nss\
(Figure~\ref{fig:orbit}).


\section{Conclusions}

We observed \nss\ with \xmm\ seven times during the course of a 40 day interval
in an effort to measure the local pulse frequency and frequency derivative with
high precision.  Due to unanticipated phase noise, we were unable to
phase-connect the full data set.  From systematic pulse ephemeris searches, we
identified a number of possible timing solutions, none of which had spin-down
rates close to the average spin-down rate of \nss\ over the last 5 years.   The
interpretation of the erratic long-term timing behavior of \nss\ in terms of
the usual pulsar timing noise would require timing noise levels much higher
than seen in isolated neutron stars with comparable spin-down rates.  The most
plausible explanation for the erratic long-term pulse frequency evolution of
\nss\ is the presence of a binary companion, although the current data do not
allow us to place strong constraints on system parameters.  Further efforts at
phase-coherent pulse timing observations of \nss\ are required to ($i$)
unambiguously identify the nature of the long-term pulse frequency variations,
and ($ii$) confirm and further investigate the observed short-term phase
noise.  Just one additional 40-day observing sequence with higher sampling
would yield far better constraints on orbital parameters should \nss, in fact,
possess a binary companion.

\begin{acknowledgements}
This work was supported by NASA grants NNG05GN91G and NAG5-10865.
VEZ is supported by a NASA Research Associateship Award
at NASA Marshall Space Flight Center.
\end{acknowledgements}


\begin{thebibliography}{3}

\bibitem{Ref1}
Arzoumanian Z., Nice D.J., Taylor J.H., et al., ApJ, {\bf 422},
671 (1994)

\bibitem{Ref2}
Bignami G.F., Caraveo P.A., De Luca A., et al., Nature, {\bf 423},
725 (2003)

\bibitem{Ref3}
Buccheri R., Bennett K., Bignami G.F., et al., A\&A, {\bf 128}, 245 (1983)

\bibitem{Ref4}
Cordes, J.M. \& Helfand, D.J. ApJ, {\bf 239}, 640 (1980)

\bibitem{Ref5}
Fesen R.A., Pavlov, G.G., Sanwal, D., ApJ, {\bf 636}, 848 (2006)

\bibitem{Ref6}
Gavriil F.P., Kaspi V.M., ApJL, {\bf 609}, 67 (2004)

\bibitem{Ref7}
Hailey C.J., Mori K., ApJL, {\bf 578}, 133 (2002)

\bibitem{Ref8}
Heyl J.S., Hernquist L., MNRAS, {\bf 340}, L37 (1999)

\bibitem{Ref9}
Kaplan D.L, van Kerkwijk M.H., ApJL, {\bf 635}, 65 (2005a)

\bibitem{Ref10}
Kaplan D.L, van Kerkwijk M.H., ApJL, {\bf 628}, 45 (2005b)

\bibitem{Ref11}
Mattox, J.R., Halpern, J.P., Caraveo, P.A., A\&AS, {\bf 120}, 77 (1996)

\bibitem{Ref12}
Mereghetti S., De Luca A., Caraveo P.A., et al., ApJ, {\bf 581}, 1280 (2002)

\bibitem{Ref13}
Mori, K., \& Hailey, C., astro-ph/0301161, submitted to ApJ (2006) 

\bibitem{Ref14}
Mori K., Chonco J.C., Hailey C.J., ApJ, {\bf 631}, 1082 (2005)

\bibitem{Ref15}
Pavlov, G.G., Bezchastnov, V.G.. ApJL, {\bf 635}, L61 (2005)

\bibitem{Ref16}
Pavlov, G.G., Sanwal, D., Garmire, G.P.,et al. 
In: Neutron Stars in Supernova Remnants, ed.\ P.O.\ Slane \&
B.M.\ Gaensler,  ASP Conf.\ Ser. {\bf 271}, 247 (2002a)

\bibitem{Ref17}
Pavlov, G.G., Zavlin V.E., Sanwal D., et al., ApJL, {\bf 569}, 95 (2002b)

\bibitem{Ref18}
Pavlov G.G., Sanwal D., Teter M.A., In: IAU Symp.\ 218,
Young Neutron Stars and Their Enviroments, ed. F. Camilo \&
B.M. Gaensler (San Francisco: ASP), p.~239 (2004)

\bibitem{Ref19}
Sanwal D., Pavlov G.G., Zavlin V.E., et al., ApJL, {\bf 574}, 61 (2002)

\bibitem{Ref20}
Wang Z., 
Kaplan D.L., Chakrabarty D., ApJ, submitted (astro-ph/0606686)

\bibitem{Ref21}
Woods P.M., Kaspi, V.M., Thompson, C., et al., ApJ, {\bf 605}, 378, (2004)

\bibitem{Ref22}
Woods P.M., Kouveliotou C., G\"og\"us E., et al., ApJ, {\bf 576}, 381
(2002)

\bibitem{Ref23}
Woods P.M., Kouveliotou C., Finger M.H., et al., ApJ, in press
(2006; astro-ph/0602531)

\bibitem{Ref24}
Zavlin V.E., Pavlov G.G., Sanwal D., et al., ApJL, {\bf 540}, 25 (2000)

\bibitem{Ref25}
Zavlin V.E., Pavlov G.G., Sanwal D., ApJ, {\bf 606}, 444 (2004)

\bibitem{Ref26}
Zavlin V.E., Tr\"umper, J., \& Pavlov G.G. ApJL, {\bf 525}, 959 (1999)




\end{thebibliography}
\end{document}